\documentclass[11pt,a4paper,english,nofootinbib,,superscriptaddress]{revtex4-2}
\usepackage{lmodern}

\usepackage[T1]{fontenc}
\usepackage[latin9]{inputenc}
\usepackage{babel}

\usepackage{xcolor}
\usepackage{amsmath}
\usepackage{graphicx}
\usepackage{amssymb}
\usepackage{esint}
\usepackage[unicode=true, pdfusetitle,
 bookmarks=true,bookmarksnumbered=false,bookmarksopen=false,
 breaklinks=false,pdfborder={0 0 1},backref=false,colorlinks=false]
 {hyperref}
\def\b{\begin{equation}}
\def\e{\end{equation}}

\makeatletter
\@ifundefined{textcolor}{}
{%
 \definecolor{BLACK}{gray}{0}
 \definecolor{WHITE}{gray}{1}
 \definecolor{RED}{rgb}{1,0,0}
 \definecolor{GREEN}{rgb}{0,1,0}
 \definecolor{BLUE}{rgb}{0,0,1}
 \definecolor{CYAN}{cmyk}{1,0,0,0}
 \definecolor{MAGENTA}{cmyk}{0,1,0,0}
 \definecolor{YELLOW}{cmyk}{0,0,1,0}
 }

\usepackage{latexsym}\usepackage{bm}

\makeatother

\makeatother

\begin{document} 

\title{New Chiral Generalized Minimal Massive Gravity}

\author{Mohammad Reza Setare}

\email{rezakord@ipm.ir}

\affiliation{Department of Science, Campus of Bijar, University of Kurdistan, Bijar, Iran.}

\author{Seyed Naseh Sajadi}

\email{naseh.sajadi@gmail.com}

\affiliation{Department of Science, Campus of Bijar, University of Kurdistan, Bijar, Iran.}

\author{Suat Dengiz}

\email{sdengiz@thk.edu.tr}

\affiliation{Department of Mechanical Engineering, University of Turkish Aeronautical Association, 06790 Ankara, Turkey}

\author{Ercan Kilicarslan}

\email{ercan.kilicarslan@usak.edu.tr}

\affiliation{Department of Mathematics, Usak University, 64200, Usak, Turkey.}

\date{\today}
\begin{abstract}
  
We study the generalized minimal massive gravity (GMMG) in Compere, Song and Strominger (CSS) boundary conditions employing a semi-product of a Virasoro and a $\hat{u}(1)$ Kac-Moody current algebras as the asymptotic symmetry algebra. We calculate the entropy of BTZ black holes via the degeneracy of states belonging to a Warped-CFT. We compute the linearized energy excitations by using the representations of the algebra $\hat{u}(1) \times SL(2,R)_R$ and show that energies of excitations are non-negative at (two) chiral points in the parameter space. At these special points, the charge algebra is described by either Virasoro algebra or Kac-Moody algebra. We also consider some special limits of the GMMG theory which correspond to $2+1$-dimensional massive gravity theories such as new massive and minimal massive gravity theories.


\end{abstract}
\maketitle
\section{Introduction}
The lack of a complete quantum gravity has led to extensively study lower dimensional toy models to reveal the nature of such a ultimate theory further. In this course, the Topologically Massive Gravity (TMG), by Deser, Jackiw and Templeton, is undoubtedly one of the most prominent approaches \cite{TMG1, TMG2}. This is because of the fact that TMG is renormalizable $3D$-dimensional gravity model propagating with a local dynamical massive graviton with a single helicity. Holographically, the cosmological TMG (CTMG) comprises $2+1$-dimensional anti de-Sitter (AdS) solutions corresponding to a $2$-dimensional conformal field theory (CFT) possessing two copies of Virasoro algebra with the central charges $c_{L,R} = 3l/2G(\sigma \pm 1/\mu l)$, where $\mu$ and $l$ are the Chern-Simons coupling and AdS radius, respectively.\footnote{Here, $\lambda = -1/l^2$, and $L(R)$ represents the left(right)-moving central charge, whereas $k$ stands for the topological mass parameter defined in \cite{TMG1, TMG2}.} Despite of all those inventions, the model involves some well-known shortcomings which have intensely enforced researches to deepen their works even further in all aspects to get over the presence ambiguities in the model. This includes such as the lack of a stable vacuum state or the unitary problem due to the near-boundary log-modes at the chiral point, which was first discovered in \cite{GrumillerLOG} and later approved in \cite{MaloneyStrominer} together with numerous other papers. (For this, particularly, see \cite{GrumillerREVIEWLOG} and the references therein for a comprehensive review of a complete historical account and a list of the various checks of AdS/log CFT performed in the first half decade after its discovery. In this respect, see also \cite{Mvondo-She} for a rather recent accurate historical account of this issue). The chiral limit of CTMG \cite{Chiralgravity} has a notable importance towards a well-behaved $2+1$-dimensional quantum gravity model since it specifically resolves the long-lived clash between the positivity of boundary central charges and the energy of BTZ black hole solutions \cite{BTZ}. That is, this controversy drops due to the fact that one of (left or right-moving) two central charges associated to the boundary CFT dies out. The evaporation of either left or right-moving central charge, however, results in a logarithmic excitations yielding a non-unitary boundary CFT. To particularly get over this strict flaw arising in the holographic analysis, an appealing modification of CTMG via deformation of the model by appropriate higher curvature terms has been put forward in \cite{GMMG}. The model is called Generalized Minimal Massive Gravity (GMMG) which is a proper enhancement of Minimal Massive Gravity (MMG) \cite{MMG}, and steers clear of the conflict in the unitarity of bulk and boundary. Also, it has been shown that GMMG is free from the so-called Boulware-Deser ghosts \cite{BoulwareDeserghost}-which in general emerge in massive gravity theories as extra degrees of freedom (dofs)-and has two local dynamical dofs. (See \cite{SetareHamedGMMG} for the computation of conserved charges in GMMG. Also, as another interesting completely different modification of TMG in the Standard model gauge theory perspective, see \cite{WeylTMG} whose asymptotic symmetry structure has recently been studied in \cite{LWWeylTMG}.)

Recall that the boundary conditions are extremely pivotal in finding asymptotic symmetry structure of any theory entirely. According to various recent peculiar studies which go beyond the usual Brown-Henneaux boundary conditions \cite{BrownHenneaux} and consider different viable possibilities \cite{BeyonBH1}-\cite{BeyonBH8}, there is in fact an unignorable dominant idea that $2$-dimensional CFT may not be the boundary theory of $2+1$-dimensional bare $AdS$ space at all. Here, the boundary conditions introduced by Compere, Song and Strominger (CSS) seem to be rather appealing \cite{BeyonBH2}. More precisely, CSS have demonstrated that as one considers a family of specific alternative boundary conditions, the asymptotic symmetry structure of a $2+1$-dimensional theory turns out to consist of a semidirect product of a Virasoro and $u(1)$ Kac-Moody algebras which are symmetries of the $2$-dimensional Warped-CFT's \cite{BeyonBH4, Detournay2}. (See \cite{WADS1}-\cite{ WADS12} and references therein for related some studies on Warped-CFT's.) Accordingly,  Ciambelli,  Detournay and Somerhausen have recently demonstrated that as one imposes the CSS boundary conditions to TMG, one arrives at two critical points among the existing couplings with which one gets a chiral Virasoro algebra or a $u(1)$ Kac-Moody algebra as the asymptotic symmetry \cite{NewChiralGravity}. In this work, we analyze the GMMG under the CSS boundary conditions as is done for TMG in \cite{NewChiralGravity}. We obtain the entropy of BTZ black holes by counting the degeneracy of states associated with a Warped-CFT. We compute the linearized energy excitations and demonstrate that energies of excitations are non-negative at (two) critical points in the parameter space where the charge algebra turns out to be a Virasoro  algebra or Kac-Moody algebra. We also consider some special limits of GMMG corresponding to $2+1$-dimensional massive gravity theories such as MMG and NMG \cite{MMG, NMG}.

The lay-out of paper is as follows: In Sec. \ref{sectwo}, we study the charge algebra of GMMG under the CSS boundary conditions. Here, the entropy of BTZ black holes are also computed via the degeneracy of states belonging to a Warped-CFT. In Sec. \ref{secthree}, we obtain the energy of linearized gravitons in AdS background. Sec. \ref{secconc} is dedicated to our conclusions and discussion on possible future directions. Finally, the NMG under the CSS boundary conditions is studied in the Appendix. 

\section{GMMG Under CSS Boundary Conditions}\label{sectwo}
The GMMG is constructed by generalizing the generalized massive gravity via the addition of appropriate higher curvature terms. The Lagrangian of GMMG is represented in the compact form as follows \cite{GMMG}
\begin{equation}\label{eqlag}
{\cal L}_{GMMG}=-\varsigma e.R+\dfrac{\Lambda_{0}}{6}e.e\times e+h.T(\omega)+\dfrac{1}{2\mu}\left(\omega.d\omega+\dfrac{1}{3}\omega.\omega\times \omega \right)-\dfrac{1}{m^2}\left(f.R+\dfrac{1}{2}e.f\times f\right)+\dfrac{\vartheta}{2}e.h\times h.
\end{equation}
Here, $m$ is a mass parameter of NMG term \cite{NMG}, $h$ and $f$ are auxiliary one-form fields, $\Lambda_{0}$ is the bare cosmological parameter with dimension of mass squared, $\varsigma$ denotes $\pm$ signs, $\mu$ stands for the topological mass parameter of Chern-Simons term, $\vartheta$ is a dimensionless parameter, $e$ represents dreibein, $\omega$ is a dualized spin-connection, $T(\omega)$ and $R(\omega)$ are a Lorentz covariant torsion and a curvature 2-form, respectively. As is mentioned, the equation for metric is obtained by generalizing field equation of MMG. To be more precise, let us recall that the field equation of GMMG is defined as follows \cite{GMMG}:
\begin{equation}
\bar{\sigma}G_{\mu \nu}+\Lambda_{0}g_{\mu \nu}+\dfrac{1}{\mu}C_{\mu \nu}+\dfrac{\gamma}{\mu^{2}}J_{\mu \nu}+\dfrac{s}{2m^{2}}K_{\mu \nu}= 0 \, ,
\label{eq1}
\end{equation}
where the explicit form of terms in (\ref{eq1}) respectively read 
\begin{align}\label{eq2}
&C_{\mu \nu}=\dfrac{1}{2}\epsilon_{\mu}^{\ \alpha \beta}\nabla_{\alpha}(R_{\beta \nu}-\dfrac{1}{4}g_{\nu \beta}R),\hspace{0.1cm}
J_{\mu \nu}=R_{\mu \alpha}R^{\alpha}_{\nu}-\dfrac{3}{4}R R_{\mu \nu}-\dfrac{1}{2}g_{\mu \nu}\left(R^{\alpha \beta}R_{\alpha \beta}-\dfrac{5}{8}R^{2}\right),\nonumber\\
K_{\mu \nu}=&-\dfrac{1}{2}\nabla^{2}R g_{\mu \nu}-\dfrac{1}{2}\nabla_{\mu}\nabla_{\nu} R+2\nabla^{2}R_{\mu \nu}+4 R_{m a n b}R^{a b}-
\dfrac{3}{2}R R_{\mu \nu}-R_{\alpha \beta}R^{\alpha \beta}g_{\mu \nu}+\dfrac{3}{8}R^{2}g_{\mu \nu}.
\end{align}
where $G_{\mu \nu}$ is the Einstein tensor. Note in (\ref{eq1}) that the parameter $s$ denotes sign, while the parameters $\gamma$, $\bar{\sigma}$ and $\Lambda_{0}$  are the ones that are described in terms of cosmological constant $\Lambda$, $m$, $\mu$, and the sign of Einstein-Hilbert term. One should also observe that the symmetric tensors $J_{\mu \nu}$ and $K_{\mu \nu}$ respectively originate from MMG and NMG parts. As is mentioned above, our main focus is to study GMMG in the CSS boundary conditions \cite{BeyonBH2} rather than that of the usual Brown and Henneaux as in \cite{NewChiralGravity}. For this purpose, let us first recall that the CSS boundary conditions on the metric components are described as  
\begin{align}\label{eqbound}
g_{rr}=&\dfrac{l^2}{r^2}+\mathcal{O}\left(\dfrac{1}{r^4}\right),\;\;\;g_{+ -}=-\dfrac{l^2 r^2}{2}+\mathcal{O}\left(1\right)\nonumber,\\
g_{r \pm}=&\mathcal{O}\left(\dfrac{1}{r^3}\right),\;\;\; g_{+ +}=\partial_{+}\bar{P}(x^{+})l^2 r^2+\mathcal{O}\left(1\right)\nonumber,\\
g_{- -}=&4Gl \Delta +\mathcal{O}\left(\dfrac{1}{r}\right).
\end{align}
The general solution obeying the boundary conditions can be written as
\begin{align}\label{eqmetric}
ds^2=&\dfrac{l^2}{r^2}dr^2-r^2 dx^{+}(dx^{-}-\partial_{+}\bar{P}dx^{+})+4Gl\left[\bar{L}dx^{+2}+\Delta(dx^{-}-\partial_{+}\bar{P}dx^{+})^2 \right]\nonumber\\
&\;\;\;\;\;\;\;\;\;\;\;\;\;\;\;\;\;-\dfrac{16G^2 l^2}{r^2}\Delta \bar{L} dx^{+}(dx^{-}-\partial_{+}\bar{P}dx^{+}),
\end{align}
where $l$ stands for the $AdS$ radius, $G$ is the so-called Newton's constant, $\bar{L}(x^+)$ and $\partial_+ \bar{P}(x^+)$ are dimensionless periodic chiral functions, $\Delta$ is any constant. Here, $x^{\pm}=\frac{t}{l}\pm \phi$ where $\phi \sim \phi +2 \pi $ and the conformal boundary corresponds to the limit as $\rho \rightarrow \infty$ \cite{BeyonBH2, NewChiralGravity}. Notice that the Cotton tensor of the spacetime in (\ref{eqmetric}) vanishes since it is conformally flat.

As for the GMMG, one can show that the metric in (\ref{eqmetric}) is also a solution to the GMMG field equations in (\ref{eq1}) provided that
\begin{equation}
\Lambda_{0}=\bar{\sigma} \Lambda - \Big(\dfrac{\gamma}{4\mu^2}-\dfrac{s}{4 m^2}\Big) \Lambda^2,
\end{equation}
yielding
\begin{equation}
\Lambda=\frac{\Big[\bar{\sigma} \pm \sqrt{\bar{\sigma}^2-\Lambda_0 \Big(\frac{\gamma}{\mu^2}-\frac{s}{m^2})} \Big]}{\frac{1}{2} \Big(\frac{\gamma}{\mu^2}-\frac{s}{m^2} \Big)}.
\end{equation}
Moreover, by solving the Killing equation, one gets the following Killing vectors for the metric components in (\ref{eqbound})
\begin{align}\label{killvec}
\xi =&\epsilon \partial_{+}+\left(\bar{\sigma} +\dfrac{l^2}{2r^2}\partial^{2}_{+}\epsilon \right)\partial_{-}-\dfrac{r}{2}\partial_{+}\epsilon \partial_{r}+\mathcal{O}\left(\dfrac{l^4}{r^4}\right).
\end{align}
Correspondingly, the conserved charges associated to the Killing vectors (\ref{killvec}) in the limit $r \rightarrow \infty$ as defined in \cite{SetareHamedGMMG, Barnich:2001jy} can be directly integrated on the phase space as follows
\begin{align}\label{eqcch}
Q_{\epsilon=e^{i m x^{+}}}=&\dfrac{1}{2\pi}\int_{0}^{2\pi} d\phi \; e^{i m x^{+}}\left[\left(\bar{\sigma} + \dfrac{1}{\mu l}+\dfrac{s}{4 m^2 l^2}+\dfrac{\gamma}{4\mu^2 l^2}\right)\bar{L}-\left(\bar{\sigma}-\dfrac{1}{\mu l}+\dfrac{s}{4 m^2 l^2}+\dfrac{\gamma}{4\mu^2 l^2}\right)\Delta (\partial_{+}\bar{P})^{2}\right],\nonumber\\
Q_{\sigma=e^{i m x^{+}}}=&\dfrac{1}{2\pi}\int_{0}^{2\pi} d\phi \; e^{i m x^{+}} \left(\bar{\sigma} -\dfrac{1}{\mu l}+\dfrac{s}{4 m^2 l^2}+\dfrac{\gamma}{4\mu^2 l^2}\right)\left(\Delta+ 2\Delta \partial_{+}\bar{P}\right).
\end{align}
Furthermore, one can show that the asymptotic symmetry generators which can be represented as
\begin{equation}
L_{n}=Q\left(\epsilon=e^{i n x^{+}}\right),\;\;\;\;\;\;\;\;\;\;\;M_{n}= Q\left(\sigma=e^{i n x^{+}}\right),
\end{equation}
comply with the following algebra
\begin{align}\label{algebra}
i\left[L_{m},L_{n}\right]=&(m-n)L_{m+n}+\dfrac{c_{R}}{12}m^3\delta_{n+m,0},\nonumber\\
i\left[L_{m},M_{n}\right]=&-mM_{m+n},\nonumber\\
i\left[M_{m},M_{n}\right]=&\dfrac{k_{KM}}{2}m\delta_{n+m,0},
\end{align} 
where the charges are given as follows
\begin{equation}\label{eqq11}
c_{R}=\dfrac{3 l}{2G}\left(\bar{\sigma} +\dfrac{1}{\mu l}+\dfrac{s}{4 m^2 l^2}+\dfrac{\gamma}{ 4\mu^2 l^2}\right),\;\;\;\;\;\;\;\;\;\;\; k_{KM}=-4\left(\bar{\sigma}+\dfrac{s}{4m^2 l^2}-\dfrac{1}{\mu l}+\dfrac{\gamma}{4\mu^2 l^2}\right)\Delta.
\end{equation}
Observe that the commutators are that of a Virasoro-Kac-Moody algebra as in \cite{NewChiralGravity}. As an explicit example, let us now compute the entropy of the BTZ black hole via counting the degeneracy of states in the dual $2$-dimensional Warped-CFT. In this regard, let us first note that the rotating BTZ metric is 
\begin{equation}
ds^2=-f(r)dt^2+\dfrac{dr^2}{f(r)}+r^2(Ndt+d\phi)^2,
\end{equation}
where the existing functions are 
\begin{equation}
f(r)=\dfrac{r^2}{l^2}-8GM+\dfrac{16G^2 J^2}{r^2},\;\;\;\; N=-\dfrac{4GJ}{r^2}.
\end{equation}
As is well-known, the black hole horizons are located at the following radii 
\begin{equation}
r_{\pm}=\sqrt{2Gl(lM+J)}\pm\sqrt{2Gl(lM-J)}.
\end{equation}
Notice that the BTZ entropy in GMMG has been computed in \cite{SetareHamedGMMG} as follows
\begin{equation}
S=4\pi\left[\left(\bar{\sigma}+\dfrac{\gamma}{2\mu^2 l^2}+\dfrac{s}{2m^2 l^2}\right)r_{+}-\dfrac{r_{-}}{\mu l}\right].
\end{equation}
We expect this to be reproduced by counting the degeneracy of states in the dual Warped-CFT. The warped Cardy formula takes the form
\begin{equation}\label{eqentropy}
S_{WCFT}=4\pi \sqrt{-M_{0}M_{0vac}}+4\pi \sqrt{-L_{0}L_{0vac}}.
\end{equation}
Note that, in this expression, the subscript $vac$ refers to the charges of the vacuum, $M = -1/8G$ and $J = 0$ for vacuum. For the BTZ black hole, one gets the zero modes as follows
\begin{align}
L_{0}=&\left(\bar{\sigma} +\dfrac{1}{\mu l}+\dfrac{\gamma}{4\mu^2 l^2}+\dfrac{s}{4m^2 l^2}\right)\left(\dfrac{ lM-J}{2}\right),\nonumber\\
M_{0}=&\left(\bar{\sigma} -\dfrac{1}{\mu l}+\dfrac{\gamma}{4\mu^2 l^2}+\dfrac{s}{4m^2 l^2}\right)\left(\dfrac{lM+J}{2}\right),\end{align}
which, for the vacuum, reduce to  
\begin{align}
M_{0vac}=-\dfrac{\left(\bar{\sigma} -\dfrac{1}{\mu l}+\dfrac{\gamma}{4\mu^2 l^2}+\dfrac{s}{4m^2 l^2}\right)l}{16G}\nonumber, \,\,\,\,\,\,\,\, L_{0vac}=-\dfrac{\left(\bar{\sigma} +\dfrac{1}{\mu l}+\dfrac{\gamma}{4\mu^2 l^2}+\dfrac{s}{4m^2 l^2}\right)l}{16G}.
\end{align}
Plugging these in (\ref{eqentropy}) one finds that $S =S_{WCFT}$. Furthermore, the energy of BTZ black hole turns out to be as follows \cite{SetareHamedGMMG}
 \begin{equation}
 E=\left(\bar{\sigma} +\dfrac{\gamma}{2\mu^2 l^2}+\dfrac{s}{2 m^2 l^2}\right)\dfrac{r_{+}^{2}+r_{-}^{2}}{l^2}-\dfrac{2r_{+}r_{-}}{\mu l^3},
 \end{equation}
 which, at the critical points where $(k_{KM}=0, c_{R}=0)$\footnote{Note that either the Virasoro algebra or the Kac-Moody algebra vanish at the particular points $ \mu l\left(\bar{\sigma}+\dfrac{\gamma}{4\mu^{2}l^2}+\dfrac{s}{4m^2 l^2}\right)=\pm 1$.}, become
 \begin{align}
 E=&\left(\dfrac{2}{\mu l^3}(r_{+}^2+r_{-}^2-r_{+}r_{-})-\dfrac{\bar{\sigma}}{l^2}(r_{+}^2+r_{-}^2)\right), \;\;\;\ \text{at $k_{KM}=0$}\\
  E=&-\left(\dfrac{2}{\mu l^3}(r_{+}^2+r_{-}^2+r_{+}r_{-})+\dfrac{\bar{\sigma}}{l^2}(r_{+}^2+r_{-}^2)\right), \;\;\;\ \text{at $c_{R}=0$}.
 \end{align}
 
\section{The energy of gravitons}\label{secthree}
In this part, we will obtain the energy of the linearized gravitons in global AdS background. To this end, we will consider the following $2+1$-dimensional AdS spacetime in global coordinates 
\begin{equation}
ds^{2}=-\dfrac{l^2}{4}\left[-4d\rho^2 +dx^{+2}+2\cosh(2\rho)dx^{+}dx^{-}+dx^{-2}\right].
\end{equation}
By defining the linearized excitations around the AdS background metric as 
\begin{equation}
g_{\mu \nu}=\bar{g}_{\mu \nu}+h_{\mu \nu},
\end{equation}
wherein $\bar{g}_{\mu\nu}$ and $h_{\mu \nu}$ respectively are the background metric and an adequately small perturbation, one gets the linearized equations of motion belonging to GMMG as follows \cite{GMMG}
\begin{equation}\label{eqom}
\bar{\sigma} {\cal G}_{\mu \nu}^{(L)}+\Lambda_{0}h_{\mu \nu}+\dfrac{1}{\mu}{\cal C}_{\mu \nu}^{(L)}+\dfrac{\gamma}{\mu^{2}}{\cal J}_{\mu \nu}^{(L)}+\dfrac{s}{2 m^2}{\cal K}_{\mu \nu}^{(L)}=0,
\end{equation}
where $L$ represents linearized. Here, the linearized tensors are 
\begin{align}\label{eqc}
 {\cal G}^{(L) \mu \nu}=& R^{(L) \mu \nu}-\dfrac{1}{2}g^{\mu \nu} R^{(L)}-2 \Lambda h^{\mu \nu},\;\;\;
{\cal C}^{(L) \mu \nu}=\dfrac{1}{\sqrt{-\bar{g}}}\epsilon^{\mu \alpha \beta}\bar{g}_{\beta \sigma}\bar{\nabla}_{\alpha}\left(\mathcal{R}^{(L)\sigma \nu}-\dfrac{1}{4}\bar{g}^{\sigma \nu}\mathcal{R}^{(L)}+2\Lambda h^{\sigma \nu}\right)\nonumber\\
{\cal K}^{(L)}_{\mu \nu}=&2\bar{\Box}\mathcal{G}^{(L)\mu \nu}+\dfrac{1}{2}\bar{g}^{\mu \nu}\bar{\Box}\bar{R}^{(L)}-\dfrac{1}{2}\bar{\nabla}^{\mu}\bar{\nabla}^{\nu}\mathcal{R}^{(L)}-5\Lambda \mathcal{G}^{(L)\mu \nu}-\Lambda \bar{g}^{\mu \nu}\bar{R}^{(L)}+\dfrac{1}{2}\Lambda^{2}h^{\mu \nu}\nonumber\\
{\cal J}^{(L)}_{\mu \nu}=&-\dfrac{1}{2}\Lambda {\cal G}^{(L)}_{\mu \nu}-\dfrac{1}{4}\Lambda^{2}h{\mu \nu},
\end{align}
with
\begin{align}
\mathcal{R}^{(L}_{\mu \nu} =\dfrac{1}{2}\left[-\bar{\nabla}^{2}h_{\mu \nu}-\bar{\nabla}_{\mu}\bar{\nabla}_{\nu}h +\bar{\nabla}_{\mu}\bar{\nabla}_{\sigma}h^{\sigma}_{\nu}+\bar{\nabla}
_{\nu}\bar{\nabla}_{\sigma}h^{\sigma}_{\mu}\right],\;\;\; 
\mathcal{R}^{(L)} =-\bar{\nabla}^{2}h+\bar{\nabla}_{\rho}\bar{\nabla}_{\sigma}h^{\rho \sigma}-2\Lambda h. 
\end{align}
Moreover, in transverse and traceless gauge
\begin{equation}
\bar{\nabla}_{\mu}h^{\mu \nu}=h=0,
\end{equation}
together with the definition of similar mutually orthonormal operators as in \cite{Chiralgravity}, the equations of motion turn out to be \cite{GMMG}
\begin{equation}\label{eqom1}
\Big(\bar{\nabla}^{2}+\dfrac{2}{l^2}\Big)\Big[h_{\mu \nu}+\dfrac{s\tilde{m}^2}{\tilde{\mu}}\epsilon_{\mu}^{\alpha \beta}\bar{\nabla}_{\alpha}h_{\beta \nu}+\Big(s\tilde{m}^{2}+\dfrac{5}{2 l^2}\Big)h_{\mu \nu}\Big]=0,
\end{equation}
where the relevant parameters are defined respectively as follows 
\begin{equation}
\tilde{m}^2=\dfrac{\tilde{\mu}}{\mu}m^{2},\;\;\;\;\;\; \tilde{\mu}=\bar{\sigma}\mu+\dfrac{\gamma}{2\mu l^2}.
\end{equation}
As is substantiated in \cite{NewChiralGravity}, the solution to the equations of motion in (\ref{eqom1}) takes the following structure
\begin{equation}\label{eqpert}
h_{\mu \nu}=e^{-i(h x^{+}+px^{-})}f_{\mu \nu},
\end{equation}
where $h$ and $p$ are weight of primary states as
\begin{equation}
\mathcal{L}_{0}\vert h_{\mu \nu}\rangle = h \vert h_{\mu \nu}\rangle , \;\;\;\;\; \mathcal{P}_{0}\vert h_{\mu \nu}\rangle = p \vert h_{\mu \nu}\rangle,
\end{equation}
with the operators
\begin{equation}
\mathcal{L}_{0}=i\partial _{+},\;\;\;\;\;\; \mathcal{P}_{0}=i \partial_{-}.
\end{equation}
By using of the transverse, traceless, and highest-weight condition, one can obtain $f_{\mu \nu}$, whose components will depend on $p$, $h$ and integration constants $\alpha$ and $\beta$. So, the components of $f_{\mu \nu}$ in the Fefferman-Graham coordinates are given as \cite{NewChiralGravity}
\begin{align}\label{eqpert1}
f_{++}=&\dfrac{1}{4}\cosh^{4-2H}\rho \tanh^{P-H}\rho (4\beta \tanh^{2}\rho +\alpha \tanh^{4}\rho)\nonumber,\\
f_{+-}=&\dfrac{1}{2}\cosh^{2(1-H)}\rho \tanh^{P-H}\rho (\beta \tanh^{2}\rho)\nonumber,\\
f_{+\rho}=&\dfrac{i}{32}\sinh^{-1}\rho \cosh^{-(1+2H)}\rho \tanh^{P-H}\rho (4(2\beta -\alpha)\cosh 2\rho -8\beta +3\alpha +\alpha \cosh 4\rho)\nonumber\\
f_{--}=&0\nonumber,\\
f_{-\rho}=&-\dfrac{i}{4}\cosh^{-1}\rho \sinh^{-1}\rho \sinh^{-H}2\rho \tanh^{P-H}\rho (\sinh^{H}2\rho \cosh^{-2H}\rho (-\beta \cosh2\rho + \beta))\nonumber,\\
f_{\rho \rho}=&\sinh^{-2-H}2\rho \tanh^{P-H}\rho (\cosh^{4-2H}\rho \sinh^{H}2\rho ((4\beta -\alpha)\tanh^{4}\rho)).
\end{align}

Using all those setups, one can easily show that the energy of right photon, right and massive graviton modes $h_{\mu \nu}^{P, R, M}$ become as follows \cite{GMMG}
\begin{equation}\label{eqenergy}
E_{P,R,M}=-\dfrac{1}{4\pi G}\left(\bar{\sigma}+\dfrac{\gamma}{2\mu^{2}l^2}+\dfrac{s}{2m^2 l^2}\right)\int d^{2}x \sqrt{-g}\bar{\nabla}^{0}h_{\mu \nu}^{P,R,M}\dot{h}^{\mu \nu}_{P,R,M}.
\end{equation}
Then, by inserting equations (\ref{eqpert}) and (\ref{eqpert1}) into (\ref{eqenergy}) and regularity of $f_{\mu \nu}$, one ultimately gets 
\begin{equation}\label{massiveenergy}
E_{M}=\dfrac{\alpha^2}{8 G l^4}\left(\bar{\sigma}+\dfrac{\gamma}{2\mu^{2}l^2}+\dfrac{s}{2m^2 l^2}\right)\left[\dfrac{(p+1)^2}{2p+3}\right],
\end{equation}
for massive mode $\left(\beta=0, h=p+2, p=\frac{1}{4}\left(-2+\frac{s\tilde{m}^2 l}{\tilde{\mu}}\pm\sqrt{2-4s \tilde{m}^2 l^2+\frac{s^2 \tilde{m}^4 l^2}{\tilde{\mu}^{2}}}\right)\right)$. For the right graviton mode ($h=2, p=0, \beta=0$) and right photon mode ($p=\alpha=0, h=1$), the energies respectively become as follows 
\begin{equation}\label{photongravenergy}
E_{R}=\dfrac{\alpha^{2}}{24 G l^4}\left(\bar{\sigma}+\dfrac{\gamma}{2\mu^{2}l^2}+\dfrac{s}{2m^2 l^2}\right),\;\;\;\; E_{P}=\dfrac{3\beta^2}{G l^4}\left(\bar{\sigma}+\dfrac{\gamma}{2\mu^{2}l^2}+\dfrac{s}{2m^2 l^2}\right).
\end{equation}
Note that the left graviton mode does not exist; rather there comes a right photon mode in the CSS boundary conditions. Observe that one gets the energies of the modes of MMG in the limits as $1/m^2 \rightarrow 0$ in (\ref{massiveenergy}) and (\ref{photongravenergy}) for the CSS boundary conditions, while the limits $\mu \rightarrow \infty$ and $\gamma \rightarrow 0$ yield that of NMG whose derivation is also given in the appendix. Now, let us discuss the energies of the dynamical modes at the chiral points: first of all, for the chiral point where $k_{KM}=0$, we have
\begin{equation}
E_{M}=\dfrac{\alpha^2}{8 G l^4}\left(2-\bar{\sigma} \mu l\right)\left[\dfrac{(p+1)^2}{2p+3}\right],\;\;\;\; E_{R}=\dfrac{\alpha^2 (2-\bar{\sigma} \mu l)}{24 G\mu l^5},\;\;\;\; E_{P}=\dfrac{3\beta^2(2-\bar{\sigma} \mu l)}{G\mu l^5}.
\end{equation}
Observe that as $\bar{\sigma}< 2/\mu l$, the energies of right graviton and photon modes are positive. Moreover, the energy of the massive graviton is positive if $\bar{\sigma}< 2/\mu l$ and $p>-3/2$. Secondly, for the other chiral point where $c_{R}=0$, we get
\begin{equation}
E_{M}=-\dfrac{\alpha^2}{8 G l^4}\left(2+\bar{\sigma} \mu l\right)\left[\dfrac{(p+1)^2}{2p+3}\right],\;\;\;\; E_{R}=-\dfrac{\alpha^2 (2+\bar{\sigma} \mu l)}{24 G\mu l^5},\;\;\;\;\;\; E_{P}=-\dfrac{3\beta^2 (2+\bar{\sigma} \mu l)}{G\mu l^5}.
\end{equation}
Notice that as $\bar{\sigma}< -2/\mu l$, the energies of right graviton and photon modes are positive. Moreover, the energy of the massive graviton is positive if $\bar{\sigma}< - 2/\mu l$ and $p>-3/2$. Finally, observe that as $\bar{\sigma}=- 2/\mu l$, all the energies of the modes become zero. 

\section{Conclusion}\label{secconc}
In this paper, we have studied the GMMG in the CSS boundary conditions wherein the asymptotic symmetry group turns out to be a semi-product of a Virasoro algebra and a $\hat{u}(1)$ Kac-Moody current algebra. By making use of the representations of the algebra $\hat{u}(1)\times SL(2,R)_R$, we have calculated the linearized energy excitations. Here, we have noted that the model has intriguing properties at two special points in the parameter space: more precisely, in the first case where 
\begin{equation}\label{eq37}
\mu l\left(\bar{\sigma}+\dfrac{\gamma}{4\mu^{2}l^2}+\dfrac{s}{4m^2 l^2}\right)=+1,
\end{equation}
we only have a Virasoro algebra as the asymptotic symmetry group. In this case, the energies of right graviton and photon modes turn out to be positive if $\bar{\sigma}<2/\mu l$, while the energy of massive graviton excitation becomes positive for $\bar{\sigma}<2/\mu l$ and $p>-3/2$ and finally the energies of BTZ black holes are positive for $\sigma <0$. On the other side, for the second case where
\begin{equation}\label{eq38}
 \mu l\left(\bar{\sigma}+\dfrac{\gamma}{4\mu^{2}l^2}+\dfrac{s}{4m^2 l^2}\right)=-1,
\end{equation}
the $\hat {u}(1)$ Kac-Moody current algebra turns out to be the associated asymptotic symmetry group. Here, the energies of right graviton and photon excitations become positive for $\bar{\sigma} <-2/\mu l$, the energy of massive graviton mode is positive if $\bar{\sigma}<-2/\mu l$ and $p>-3/2$ and also the energies of BTZ black holes become positive for $\sigma <0$. Notice that the central charges (\ref{eqq11}) are different from the central charges of GMMG with Brown-Henneaux boundray conditions. So, unlike TMG \cite{NewChiralGravity} at the chiral point (\ref{eq37}) and (\ref{eq38}) for GMMG the energy of massive graviton, right-moving graviton or photon does not vanish; but rather they vanish at the points $\bar{\sigma}\mu l= \pm 2$ for both chiral points, respectively. We have also observed that one easily gets the energies of the modes of MMG in the limits as $1/m^2 \rightarrow 0$ in (\ref{massiveenergy}) and (\ref{photongravenergy}), whereas the limits $\mu \rightarrow \infty$ and $\gamma \rightarrow 0$ yield that of NMG in the context of CSS boundary conditions. These observations imply that imposing the CSS boundary conditions to GMMG may provide a legitimate $3$-dimensional gravity model in the holographic context where the dual field theory becomes $2$-dimension Warped-CFT. Now that we have shown that GMMG in the CSS boundary conditions has potential to procure a holographically legitimate $2+1$-dimensional gravity theory, one shall analyze it in all the prominent perspectives as future projects. Here, showing that the BTZ black hole solutions are the only stationary ones that have axial symmetry seems to particularly be more compelling at the first place in the parallel line with new chiral gravity \cite{NewChiralGravity}. Of course, quest of whether there emerge logarithmic solutions at these chiral points or not and if so, whether or not they bring any new information to our previous information about the chiral points in usual TMG seem to be also an indispensable future direction.

\section{Acknowledgments}
We would like to thank Daniel Grumiller for his valuable and very constructive suggestions. We also wish to thank L. Ciambelli, S. Detournay and A. Somerhausen for sharing their codes. The works of E.K. and S.D. are partially supported by the TUBITAK Grant No. 119F241.
\appendix

\section{NMG in the CSS Boundary Conditions}
Even if one can the asymptotic algebra and thus the energies of the dynamical dofs for NMG directly by taking the limits $\mu \rightarrow \infty $ and $\gamma \rightarrow 0$ in the related results for the GMMG, we shall explicitly tackle the NMG in the CSS boundary conditions to ascertain the consequences in this part. To this end, let us first remember that the Lagrangian of NMG model is \cite{NMG}
\begin{equation}\label{eqlag}
I=\dfrac{1}{16\pi G}\int d^{3}x\sqrt{-g}\left[\bar{\sigma}R-2\Lambda_0 -\dfrac{1}{m^2}\left(R_{\mu \nu}R^{\mu \nu}-\dfrac{3}{8}R^{2}\right)\right],
\end{equation}
which leads to the field equations
\begin{equation}\label{eqnmg}
\bar{\sigma}G_{\mu \nu}+\Lambda_0 g_{\mu \nu}+\dfrac{1}{2m^{2}}K_{\mu \nu}= 0 \, ,
\end{equation}
where 
\begin{align}\label{eq2}
K_{\mu \nu}=&-\dfrac{1}{2}\nabla^{2}R g_{\mu \nu}-\dfrac{1}{2}\nabla_{\mu}\nabla_{\nu} R+2\nabla^{2}R_{\mu \nu}+4 R_{m a n b}R^{a b}-
\dfrac{3}{2}R R_{\mu \nu}-R_{\alpha \beta}R^{\alpha \beta}g_{\mu \nu}+\dfrac{3}{8}R^{2}g_{\mu \nu},
\end{align}
and $G_{\mu \nu}$ is the Einstein tensor.\\
The metric (\ref{eqmetric}) is a solution to the NMG field equations provided
\begin{equation}
\Lambda_0 =\left(\bar{\sigma} +\dfrac{\Lambda}{4 m^2}\right) \Lambda.
\end{equation}
The conserved charges associated to the Killing vectors (\ref{killvec}) in the limit $r \rightarrow \infty$ as follows
\begin{align}\label{eqcch1}
Q_{\epsilon=e^{i m x^{+}}}=&\dfrac{1}{2\pi}\int_{0}^{2\pi} d\phi \; e^{i m x^{+}}\left(\bar{\sigma} +\dfrac{1}{4 m^2l^2}\right)\left(\bar{L}-\Delta (\partial_{+}\bar{P})^{2}\right),\nonumber\\
Q_{\sigma=e^{i m x^{+}}}=&\dfrac{1}{2\pi}\int_{0}^{2\pi} d\phi \; e^{i m x^{+}} \left(\bar{\sigma} +\dfrac{1}{4 m^2 l^2}\right)\left(\Delta + 2\Delta \partial_{+}\bar{P}\right).
\end{align}
The generators $L, M$ of the asymptotic symmetry group satisfy the Kac-Moody algebra (\ref{algebra}) with  
\begin{equation}
c_{R}=\dfrac{3l}{2G}\left(\bar{\sigma} +\dfrac{1}{4 m^2l^2}\right),\;\;\;\;\;\;\;\;\;\;\; k_{KM}=-4\Delta \left(\bar{\sigma}+\dfrac{1}{4m^2 l^2}\right).
\end{equation}
The BTZ metric is a solution for NMG with the entropy as 
\begin{equation}
S=4\pi \left(\bar{\sigma}+\dfrac{1}{2m^2 l^2}\right)r_{+},
\end{equation}
by using of warped Cardy formula (\ref{eqentropy}) and $M = -1/8G$, $J = 0$ for vacuum and for BTZ black hole
\begin{equation}
M_{0}=\left(\bar{\sigma}+\dfrac{1}{4m^2 l^2}\right)\left(\dfrac{lM+J}{2}\right) , \hskip 0.8 cm  L_{0}=\left(\bar{\sigma} +\dfrac{1}{4m^2 l^2}\right)\left(\dfrac{lM-J}{2}\right),
\end{equation}
and for vacuum 
\begin{equation}
M_{0vac}=-\dfrac{\left(\bar{\sigma} +\dfrac{1}{4m^2 l^2}\right)l}{16G}, \hskip 0.8 cm L_{0vac}=-\dfrac{\left(\bar{\sigma} +\dfrac{1}{4m^2 l^2}\right)l}{16G},
\end{equation}
one ultimately arrives at $S =S_{WCFT}$.

\end{document}